# Carbonic Acid Revisited: Vibrational Spectra, Energetics and the Possibility of Detecting an Elusive Molecule

Stefan E. Huber,[1,*] Silvia Dalnodar,[2] Wolfgang Kausch,[2] Stefan Kimeswenger[2] and Michael Probst[1,*]

[1] Institute of Ion Physics and Applied Physics , University of Innsbruck, Technikerstr. 25, A-6020 Innsbruck, Austria

[2] Institute of Astro- and Particle Physics, University of Innsbruck, Technikerstr. 25, A-6020 Innsbruck, Austria

[*] Corresponding authors. Electronic mail: s.huber@uibk.ac.at, michael.probst@uibk.ac.at

**Abstract.** We calculate harmonic frequencies of the three most abundant carbonic acid conformers. For this, different model chemistries are investigated with respect to their benefits and shortcomings. Based on these results we use perturbation theory to calculate anharmonic corrections at the ωB97XD/aug-cc-pVXZ, X=D, T, Q, level of theory and compare them with recent experimental data and theoretical predictions. A discrete variable representation method is used to predict the large anharmonic contributions to the frequencies of the stretching vibrations in the hydrogen bonds in the carbonic acid dimer. Moreover, we re-investigate the energetics of the formation of the carbonic acid dimer from its constituents water and carbon dioxide using a high-level extrapolation method. We find that the ωB97XD functional performs well in estimating the fundamental frequencies of the carbonic acid conformers. Concerning the reaction energetics, the accuracy of ωB97XD is even comparable to the high-level extrapolation method. We discuss possibilities to detect carbonic acid in various natural environments such as Earth's and Martian atmospheres.

## I. Introduction

Carbonic acid ($H_2CO_3$) plays a role in a variety of biological, geochemical and other processes. In the human body, for instance, it acts as an intermediate in the respiratory evacuation of $CO_2$[1] and works as an important buffer to stabilize the pH value of blood.[2] Carbonic acid appears as an intermediate in the acidification of the seawater induced by uptake of large amounts of carbon dioxide from the atmosphere.[3] Carbonate minerals as salts of carbonic acid, are important in geology[4] with calcite and magnesite being the most abundant ones. Furthermore, it is believed that carbonic acid might play a role in the earth's upper atmosphere, particularly in solid form in cirrus clouds.[5] It might also participate in outer space chemistry, where $H_2CO_3$ could be present in solar system ices that contain both water and carbon dioxide and are exposed to radiation from the solar wind and planetary magnetospheres.[5,6]

Remarkably, the very existence of carbonic acid as a discrete molecular species in the gas phase has been controversial for a long time. Until recently, the only evidence that gaseous carbonic acid exists were provided by Terlouw et al.,[7] who reported a weak mass spectrometric signal at m/z=62 Thomson, by Mori et al.,[8] who obtained microwave spectra of the carbonic acid cis-trans monomer, and by Hage et al.,[5] who recondensed solid carbonic acid from the gas phase. Very recently, however, Bernard et al. managed to measure for the first time laboratory infrared spectra of gaseous carbonic acid species trapped in a solid matrix of either neon or argon.[9] Obviously, these spectra are thought to represent a benchmark for possible identification of naturally occurring carbonic acid vapor.

Accompanying the experiments, a significant amount of theoretical and computational work has been carried out in order to investigate the chemical properties of carbonic acid and its stability. A few of them, in particular those which are important for the present work, are discussed in the following. In 1995, Wight et al.[10] investigated the potential energy surface and the harmonic vibrational spectrum of carbonic acid monomers at the MP2/6-31++G** level of theory. It was shown that the global minimum structure of the monomer is cis-cis, but that the cis-trans isomer is less stable by only 1.6 kcal/mol including a ZPE correction. The two isomers differ by rotation of one OH bond, see FIG. 1a and b. Furthermore, it has been shown that these carbonic acid monomers are only metastable with respect to dissociation to water and carbon dioxide by 10.4 kcal/mol, but reactants and products are separated by a high potential barrier of about 51.9 kcal/mol. Hence, the carbonic acid cis-cis and cis-trans monomers are kinetically stable, but thermodynamically they would decompose. Two years later, in 1997, Liedl et al.[11] studied the energetics of the carbonic acid dimer consisting of two cis-cis monomers. Such a cyclic dimer is stabilized by two strong hydrogen bonds, see FIG. 1c. They investigated whether the dimer is thermodynamically stable with respect to water and carbon dioxide. It was concluded that with the best then feasible method, i.e. MP2/aug-cc-pVTZ calculations, that 'it is not possible to judge […] if the zero-point-corrected energy for the formation of the carbonic acid dimer from water and carbon dioxide is slightly positive or slightly negative; however, it is astonishingly close to zero'.[11] Hence, in gaseous carbonic acid one can most probably expect a mixture of the cis-cis and cis-trans monomers and the carbonic acid dimer shown in FIG. 1. A study of Tossell[12] in 2006 on the $H_2CO_3$ cis-cis and cis-trans monomers (and on oligomers composed of cis-cis monomers, using the CBSB7 B3LYP model chemistry and approximations for anharmonic frequencies at the CCSD(T)/6-311G+(2d,p) level revealed the importance of going beyond the harmonic frequency approximation for a reliable calculation of infrared spectra of carbonic acid. Particularily the deviations from the harmonic values for stretching and bending of bonds that include hydrogens can be quite large and this is even more the case for hydrogen bonds. Tossell found that the oligomers are more stable the larger they are.[12] In the gas phase one can, however, expect that oligomers are mainly restricted to the dimer species, since larger ones require more collision events and are thus thought to be very limited due to probability considerations.[9] Bernard et al. calculated the harmonic frequencies at the MP2/aug-cc-pVDZ level of theory to compare with their experimental results and obtained a good agreement between theoretical harmonic frequencies and experimental fundamental ones.[9] In fact, this raises the question of the reliability of this model chemistry for predicting accurate frequencies, as including anharmonic corrections might cause a considerable redshift of the line positions. The assignment of spectral lines in the work of Bernard et al.[9] is based on isotope shifts and not on direct comparison of frequencies, i.e. the assignments of vibrational modes[9] are still thought to be well based and only the reliable prediction of infrared spectra with MP2/aug-cc-pVDZ calculations is questioned. Murillo et al.[13] investigated the energetics of different dimer geometries and especially the effect of the geometry on the strength of hydrogen bonds. They found the strongest hydrogen bonds to be associated with the dimer composed of one cis-cis and one cis-trans monomer, not for the one with two cis-cis units. The latter dimer, however, remains the energetically most stable one at the CCSD(T)/aug-cc-pVDZ//MP2/6-311++G** level of theory. In 2011, Wu et al.[14] calculated a variety of molecular properties of carbonic acid isomers at the CCSD(T)//B3LYP/aug-cc-pVTZ level of theory. They reported harmonic frequencies and investigated the potential energy surface of $H_2CO_3$. In agreement with earlier work[10] they found the cis-cis and cis-trans monomers to be the most stable ones, but lying 6.72 kcal/mol (8.28 kcal/mol for the cis-trans monomer) above water and carbon dioxide and thus being only kinetically stabilized by a barrier of about 41.58 kcal/mol (43.14 kcal/mol for the cis-trans monomer).

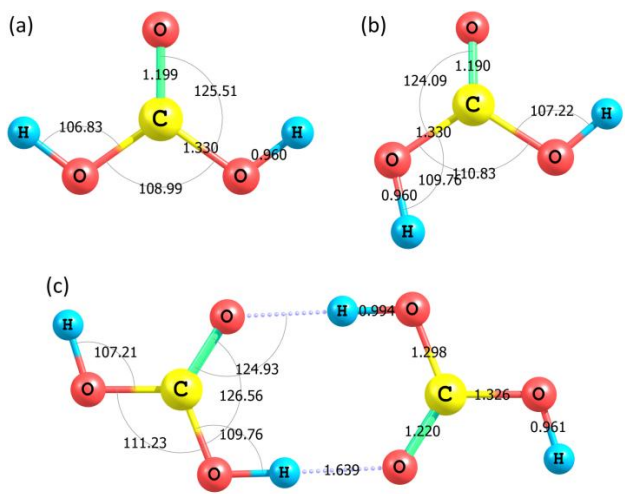

FIG. 1. Optimized geometries at the ωB97XD/aug-cc-pVQZ level of theory for (a) the cis-cis monomer, (b) the cis-trans monomer and (c) the energetically lowest dimer of carbonic acid. Bond lengths are given in Ångstrom and bond angles in degrees.

In this work we aim to improve on the aforementioned results, especially concerning the infrared absorption spectra for gas-phase carbonic acid species and the energetics of carbonic acid dimer (referring for the rest of this paper to the one composed of two cis-cis monomers). The quantum chemical methods that are used are described in section 2.

In section 3.1, we shall compare several commonly used methods in quantum chemistry to calculate harmonic frequencies. In section 3.2, we derive anharmonic corrections and discuss the basis set dependencies of the results. Concerning the energetics of the carbonic acid dimer we aim for giving a definite answer on the question whether it is lower in energy than its constituents water and carbon dioxide in section 3.3. To this aim, we restrict our calculations to the three lowest-energy carbonic acid species depicted in FIG. 1, namely the cis-cis and cis-trans monomer and the dimer composed of two cis-cis monomers.

We use these theoretical considerations for the aim to tackle a possible detection of carbonic acid in natural environments from an observational point of view as well in section 3.4. Particularly, we discuss Earth's and Martian's atmospheres as regions of naturally occurring carbonic acid and what problems and limitations one has to face aiming for a successful detection of this elusive molecule. In section 4, we summarize the work.

## II. Methodology

### A. Frequencies

We calculate harmonic frequencies with three standard methods of quantum chemistry. Two of these stem from density functional theory, namely the B3LYP functional, which uses Becke's three parameter hybrid exchange functional[15] and a correlation functional of Lee, Yang and Parr with both local and non-local terms,[16] and the ωB97XD functional of Head-Gordon et al.[17] including long range corrections as well as empirical dispersion corrections, both of which are not included in B3LYP. We also perform MP2[18] calculations on our systems. All methods require basis sets to expand the density or wave function. We investigate the influence of the finite basis set size on the resulting IR spectra with Dunning's correlation consistent polarized basis sets augmented with diffuse functions,[19] aug-cc-pVXZ, with X=D, T, Q for the B3LYP and ωB97XD functionals and X=D only (for reasons of computational cost) for the MP2 method.

Anharmonic frequencies have been calculated at the ωB97XD/aug-cc-pVXZ level of theory with X=D, T, Q for the monomeric species and X=D, T only (for the same reasons) for the carbonic acid dimer using the MP2 approach[20] as implemented in the Gaussian 09 software.[21,22]

For the very anharmonic O-H stretching modes of the hydrogen bonds stabilizing the carbonic acid dimer another approach is necessary for an accurate estimation of the frequency shifts in the asymmetric and symmetric mode of the two hydrogens in the eight-membered ring of the dimer. The full vibrational problem of two anharmonic O···H-O oscillators is solved, i.e., the time-independent Schrödinger equation in two dimensions was solved quantum-mechanically. As details of an analogous procedure can be found in an earlier work[23], we will only give a short overview in the following. In a first step a two-dimensional potential energy surface $V(q_1, q_2)$, where $q_1$ and $q_2$ denote the coordinates of the two hydrogen atoms with respect to their oxygen neighbors, i.e. the O-H bond lengths, has been calculated at the ωB97XD/aug-cc-pVXZ level of theory with X=D, T, Q. The two bond lengths were varied between 0.6 and 2.0 Å in steps of 0.1 Å. Altogether $15 \times 15 = 225$ single point calculations were performed to construct the potential energy surface for one basis set. Different basis sets have been used to determine the variation of the frequency shifts with basis set size. In a second step, a spline-type function consisting of smooth fourth-order polynomials connecting the ab initio generated grid points was calculated. In a third step, the vibrational Hamiltonian of the oscillator system was expressed as $H = K + V = -(\hbar^2/2m) \times (\partial^2/\partial q_1^2 + \partial^2/\partial q_2^2) + V(q_1, q_2)$, where $m$ denotes the mass of a hydrogen atom and $K$ the kinetic energy operator. The kinetic energy operator is expressed in matrix form via expansion in Chebyshev polynomials and subsequently transformed to a discrete variable representation (DVR). The use of Chebyshev polynomials is advantageous, as the kinetic energy matrix is diagonal in this representation. Scalar potentials are automatically diagonal in the DVR. For more detailed information regarding DVRs and their usage see e.g. the papers of Light et al. and Dawes et al.[24] Fundamental frequencies for the symmetric and asymmetric hydrogen bond stretching modes are obtained from the energy differences of the ground state and the two first excited levels, where the energies of these levels are given as the eigenvalues of the Hamiltonian. The harmonic frequencies were calculated from the curvature at the potential minimum along the two coordinates. Frequency shifts due to the anharmonicity are then determined from the difference of the harmonic and fundamental frequencies. A computer program implementing the steps discussed above has been written in MATHEMATICA.[25]

**B. Energetics**

The energetics of the carbonic acid species have been investigated by calculating ground state energies of reactants and products at various levels of theory which are then combined in a way similar to the Gx extrapolation methods[26,27] to approximate the energies related to the dissociation of the carbonic species to water and carbon dioxide. In a first step the energies have been calculated at the G4(MP2)[28] level of theory in order to have reference values. Geometries and zero-point energies were already available from the frequency calculations. In a second step, we calculated energies at the CCSD(T)/aug-cc-pVXZ//B3LYP/aug-cc-pVQZ, CCSD(T)/aug-cc-pVXZ//ωB97XD/aug-cc-pVQZ and CCSD(T)/aug-cc-pVXZ//MP2/aug-cc-pVTZ levels of theory[29] with X=D, T to investigate the effect of different optimization levels. The third step consists of the extrapolation of these energies to the CCSD(T)/aug-cc-pVQZ//(optimization level) levels of theory using Truhlar's extrapolation scheme[30]. Finally, at step four, we calculated the dissociation energy at the HF/aug-cc-pVXZ//(optimization level) level of theory with X=Q, 5 and extrapolated the resulting energies to the HF/aug-cc-pV6Z//(optimization level) level of theory using a linear two-point exponential extrapolation scheme with a global fitting parameter $\alpha = 1.63$.[31] Taking the ωB97XD/aug-cc-pVQZ level of theory as the optimization level (OL) we arrive at the following symbolic expression for the ground state energies

$$E_{ground\ state} = E\left(CCSD(T)/aug-cc-pV(D,T \xrightarrow{Truhlar})QZ//OL\right) + ZPE_{anharmonic}(OL) + \Delta_{HF}$$

where

$$\Delta_{HF} = E\left(HF/aug-cc-pV(Q,5 \xrightarrow{2-point})6Z//OL\right) - E(HF/aug-cc-pV(D,T \xrightarrow{Truhlar})QZ//OL)$$

and $ZPE_{anharmonic}$ denotes the zero-point energy approximated from the results of the anharmonic frequency calculations.[22]

All electronic structure calculations have been carried out using the Gaussian 09 software.[32]

### III. Results and Discussion

**A. Harmonic Frequencies**

We compare the harmonic frequencies obtained with B3LYP, ωB97XD and MP2 in conjunction with the aug-cc-pVXZ, with X=D, T, Q, basis sets in order to investigate dependencies on basis set size. Furthermore, we compare the results to the recently determined experimental values of Bernard et al.[9] In the latter work, experimental results are given only for a few groups of vibrational modes making no distinction between monomer and dimer species. For our purpose of a comparison of methods it is sufficient to adopt this strategy. The harmonic frequencies are given in TABLE I.

**TABLE I.** Harmonic frequencies in cm$^{-1}$ obtained from B3LYP, ωB97XD and MP2 methods in conjunction with aug-cc-pVXZ, X=D, T, Q, basis sets.

| B3LYP | | | ωB97XD | | | MP2 | | | Expt.[a] | Assign. | Molecule[b] |
|---|---|---|---|---|---|---|---|---|---|---|---|
| X=D | X=T | X=Q | X=D | X=T | X=Q | X=D | X=T | X=Q | | | |
| 3792 | 3791 | 3798 | 3874 | 3872 | 3883 | 3790 | 3804 | 3818 | 3634 | O-H symm. stretch | cis-cis |
| 3789 | 3788 | 3795 | 3871 | 3869 | 3880 | 3788 | 3802 | 3817 | 3630 | O-H asymm. stretch | cis-cis |
| 3779 | 3780 | 3786 | 3867 | 3863 | 3873 | 3783 | 3795 | 3808 | 3628 | O-H asymm. stretch | cis-trans |
| 3784 | 3791 | 3784 | 3867 | 3864 | 3875 | 3783 | - | - | 3623 | O-H symm. stretch | dimer |
| 1866 | 1871 | 1872 | 1910 | 1912 | 1915 | 1859 | 1880 | 1884 | 1836 | C=O stretch | cis-trans |
| 1770 | 1771 | 1770 | 1817 | 1815 | 1816 | 1781 | - | - | 1799 | C=O stretch | dimer |
| 1696 | 1697 | 1695 | 1746 | 1741 | 1744 | 1716 | - | - | 1727 | C=O stretch | dimer |
| 1815 | 1817 | 1818 | 1861 | 1859 | 1860 | 1814 | 1834 | 1836 | 1783 | C=O stretch | cis-cis |
| 1555 | 1552 | 1553 | 1586 | 1584 | 1583 | 1554 | - | - | 1565 | C-OH asymm stretch | dimer |
| 1521 | 1520 | 1520 | 1550 | 1550 | 1549 | 1520 | - | - | 1515 | C-OH asymm stretch | dimer |
| 1453 | 1453 | 1454 | 1488 | 1487 | 1488 | 1453 | 1467 | 1470 | 1456 | C-OH asymm stretch | cis-cis |
| 1399 | 1393 | 1394 | 1438 | 1432 | 1434 | 1400 | 1409 | 1411 | 1427 | C-OH asymm stretch | cis-trans |
| 1233 | 1223 | 1222 | 1256 | 1246 | 1246 | 1234 | - | - | 1274 | C-O-H i.p.[c] bend | dimer |
| 1150 | 1144 | 1143 | 1185 | 1177 | 1178 | 1155 | 1160 | 1160 | 1184 | C-O-H i.p.[c] bend | cis-trans |
| 1163 | 1150 | 1150 | 1192 | 1178 | 1177 | 1167 | 1166 | 1166 | 1187 | C-O-H i.p.[c] bend | cis-cis |
| 804 | 812 | 810 | 819 | 825 | 826 | 801 | - | - | 811 | CO$_3$ o.p.[c] bend[d] | dimer |
| 802 | 811 | 809 | 818 | 824 | 826 | 800 | - | - | 811 | CO$_3$ o.p.[c] bend | dimer |
| 791 | 800 | 801 | 812 | 819 | 822 | 789 | 802 | 805 | 798 | CO$_3$ o.p.[c] bend | cis-cis |
| 780 | 788 | 790 | 797 | 804 | 806 | 777 | 790 | 793 | 784 | CO$_3$ o.p.[c] bend | cis-trans |

[a] Referring to the experimental data obtained by IR spectroscopy of trapped gaseous carbonic acid in a Ne matrix, see the paper of Bernard et al.[9]
[b] Assignments to specific molecules via the molecular symmetry of the vibrational modes reported in the paper of Bernard et al.[9]
[c] i.p. and o.p. denotes in-plane and out-of-plane, respectively.
[d] CO$_3$ out-of-plane bending modes are associated with the vibration of the carbon out of the plane of the three oxygen atoms.

Except for the O-H stretching modes, the difference between double- and triple-zeta basis sets is rather large (up to about 20 cm$^{-1}$), whereas the differences between triple- and quadruple-zeta basis sets remains already much smaller (below about 5 cm$^{-1}$). Concerning the O-H stretching modes, however, increasing the basis set from triple- to quadruple-zeta can still lead to substantial changes in the resulting harmonic frequencies (up to 15 cm$^{-1}$), especially with the MP2 method. In that case, increasing the basis set size from double-zeta to quadruple-zeta changes the O-H stretching vibrations by ≈30 cm$^{-1}$, ≈25 cm$^{-1}$ in case of O=H stretching modes, ≈20 cm$^{-1}$ in case of C-OH stretching modes, ≈5 cm$^{-1}$ in case of C-O-H in-plane bending modes and ≈20 cm$^{-1}$ for the CO$_3$ out-of-plane bending modes. In the same order, the B3LYP frequencies differ by ≈7 cm$^{-1}$, ≈6 cm$^{-1}$, ≈5 cm$^{-1}$, ≈11 cm$^{-1}$ and ≈10 cm$^{-1}$ and in case of the ωB97XD functional by ≈15 cm$^{-1}$, ≈5 cm$^{-1}$, ≈6 cm$^{-1}$, ≈14 cm$^{-1}$, and ≈9 cm$^{-1}$ for the maximal shift from double-zeta to the quadruple-zeta basis set size. Thus, the deviations with respect to basis set size are smaller in case of DFT than in case of MP2. The generally weaker dependence on basis set size of DFT methods is rather well-known. Note that for most of the vibrational modes increasing the basis set size upshifts the harmonic frequencies. We conclude that at all three levels of theory, one should be careful about the accuracy of the predicted harmonic frequencies. The aug-cc-pVDZ basis set is too small, especially with MP2 calculations. Except for the O-H stretching modes, the aug-cc-pVTZ basis set yields results already close to the ones obtained with the quadruple-zeta basis. However, for the the O-H stretching modes it appears to be necessary to use even larger basis sets. In addition to these basis set dependences one observes that, except for the O-H stretching modes, the harmonic frequencies obtained with B3LYP and MP2 appear to be rather too low in absolute values. In various cases they are smaller than the experimental fundamentals. Although the general trend of increasing the basis set size is a blueshift of the harmonic frequencies, they are still very close to the experimental results, which are thought to be slightly blueshifted due to interactions between carbonic acid molecules and the surrounding neon matrix.[9] Considering from earlier work on carbonic and formic acids[12,33] that the fundamental frequencies of such systems are mostly redshifted compared to harmonic ones, we conclude from this rough inspection that the ωB97XD functional is the most reliable method amongst the three ones tested. Of course, this does not come as a surprise since it includes long-range and empirical dispersion corrections, and has been reported to perform especially well in systems exhibiting weak bonds such as hydrogen bonds.[17] For this reason, we restricted the investigation of fundamental frequencies to this functional.

**B. Anharmonic Corrections**

Harmonic and fundamental frequencies, frequency shifts and fundamental-harmonic frequency ratios obtained at the ωB97XD/aug-cc-pVXZ, X=D, T, Q levels of theory are summarized in tables 2 to 4 for the carbonic acid cis-cis monomer, the cis-trans monomer and the carbonic acid dimer, respectively. For reasons of computational cost, only double- and triple zeta basis sets were used in case of the dimer. Those vibrational modes for which experimental data is available from the paper of Bernard et al.[9] are also given in the respective tables. Tossell,[12] already calculated fundamental frequencies for the cis-cis dimer at the CCSD(T)/6-311+G(2d,p) level and corrected these for anharmonicity by adding to them the difference between B3LYP/CBSB7[34] anharmonic and harmonic frequencies, see also the work of Carbonniere et al.[35] Since these fundamental frequencies are believed to be the best available theoretical ones, we compare our results to them in TABLE III as well. Concerning the carbonic acid dimer, Tossell[12] reported only B3LYP/CBSB7 anharmonic frequencies which are included for comparison in TABLE IV.

**TABLE II.** Harmonic and fundamental frequencies and their differences (redshifts are positive) in cm$^{-1}$ and ratios of the carbonic acid cis-cis monomer at the ωB97XD/aug-cc-pVXZ, X=D,T,Q, level of theory.

| ωB97XD/aug-cc-pVDZ | | | | ωB97XD/aug-cc-pVTZ | | | | ωB97XD/aug-cc-pVQZ | | | | Expt.[a] | Vib. mode |
|---|---|---|---|---|---|---|---|---|---|---|---|---|---|
| Harm. | Fund. | Shift | Ratio | Harm. | Fund. | Shift | Ratio | Harm. | Fund. | Shift | Ratio | | |
| 3874 | 3627 | 247 | 0.936 | 3872 | 3654 | 218 | 0.944 | 3883 | 3641 | 242 | 0.938 | 3634 | O-H stretch |
| 3871 | 3624 | 247 | 0.936 | 3869 | 3651 | 218 | 0.944 | 3880 | 3638 | 242 | 0.938 | 3630 | O-H stretch |
| 1861 | 1815 | 46 | 0.975 | 1859 | 1824 | 35 | 0.981 | 1860 | 1829 | 31 | 0.984 | 1783 | C=O stretch |
| 1489 | 1445 | 44 | 0.971 | 1488 | 1448 | 40 | 0.973 | 1488 | 1452 | 36 | 0.976 | 1456 | C-OH stretch |
| 1307 | 1220 | 87 | 0.933 | 1299 | 1235 | 64 | 0.951 | 1298 | 1234 | 64 | 0.951 | | HOC i.p. bend |
| 1192 | 1110 | 82 | 0.931 | 1178 | 1118 | 60 | 0.949 | 1177 | 1117 | 60 | 0.949 | 1187 | HOC i.p. bend |
| 1001 | 981 | 20 | 0.980 | 1004 | 981 | 23 | 0.977 | 1005 | 987 | 18 | 0.982 | | C-O stretch |
| 813 | 790 | 23 | 0.972 | 819 | 800 | 19 | 0.977 | 822 | 802 | 20 | 0.976 | 798 | CO$_3$ o.p. bend |
| 611 | 347 | 266 | 0.567 | 604 | 409 | 195 | 0.677 | 604 | 396 | 208 | 0.655 | | HOC o.p. bend |
| 605 | 593 | 12 | 0.980 | 611 | 603 | 8 | 0.987 | 612 | 610 | 2 | 0.996 | | OCO bend |
| 550 | 529 | 21 | 0.961 | 555 | 541 | 14 | 0.974 | 556 | 547 | 9 | 0.983 | | OCO bend |
| 548 | 253 | 295 | 0.462 | 539 | 325 | 214 | 0.603 | 539 | 309 | 229 | 0.574 | | HOC o.p. bend |

[a] Referring to the experimental data obtained by IR spectroscopy of trapped gaseous carbonic acid in a Ne matrix, see the paper of Bernard et al.[9]

Concerning the carbonic acid cis-cis monomer, we observe a good agreement within about 5 cm$^{-1}$ between the ωB97XD/aug-cc-pVQZ results and the experiment for the O-H stretching modes at about 3630 cm$^{-1}$, the C-OH asymmetric stretch at about 1450 cm$^{-1}$ and CO$_3$ out-of-plane bending mode at about 800 cm$^{-1}$. In case of the C=O stretching and C-O-H in-plane bending modes (at about 1780 and 1190 cm$^{-1}$, respectively) the differences between theory and experiment are rather large with 46 and 70 cm$^{-1}$, respectively. There exists some basis set dependence of the frequency shifts, although most changes between T and Q are already small, except for the O-H stretch, where shifts of 247, 218 and 242 cm$^{-1}$ are obtained with the double-, triple- and quadruple-zeta basis, respectively. The trend of increasing frequencies with increasing basis set size can also be seen from both harmonics and fundamentals. Therefore, due to error cancellation, frequencies obtained with the smaller basis sets fit better to experimental ones in those cases where the ωB97XD method seems to overestimate the fundamental frequencies. Furthermore, we note that an ad-hoc inclusion of anharmonicity by empirical scaling factors might lead to significant errors, especially in case of the H-O-C out-of-plane bending modes at 300-400 cm$^{-1}$, where the ratio of fundamental to harmonic frequency is as small as 0.5-0.7 while the usual scaling factors are 0.95-0.97.[15,36] The ratios associated with the other vibrational modes are distributed in the interval 0.93-0.98.

**TABLE III.** Values like in TABLE II for the carbonic acid cis-trans monomer.

| ωB97XD/aug-cc-pVDZ | | | | ωB97XD/aug-cc-pVTZ | | | | ωB97XD/aug-cc-pVQZ | | | | Tossell[a] | Expt.[b] | Vib. mode |
|---|---|---|---|---|---|---|---|---|---|---|---|---|---|---|
| Harm. | Fund. | Shift | Ratio | Harm. | Fund. | Shift | Ratio | Harm. | Fund. | Shift | Ratio | | | |
| 3869 | 3645 | 224 | 0.942 | 3865 | 3729 | 136 | 0.965 | 3876 | 3770 | 106 | 0.973 | 3664 | | O-H stretch |
| 3867 | 3704 | 163 | 0.958 | 3863 | 3700 | 163 | 0.958 | 3873 | 3739 | 134 | 0.966 | 3660 | 3628 | O-H stretch |
| 1911 | 1866 | 45 | 0.977 | 1913 | 1870 | 43 | 0.978 | 1915 | 1873 | 42 | 0.978 | 1864 | 1836 | C=O stretch |
| 1439 | 1383 | 56 | 0.961 | 1432 | 1383 | 49 | 0.966 | 1434 | 1386 | 48 | 0.967 | 1408 | 1427 | C-OH stretch |
| 1288 | 1208 | 80 | 0.938 | 1282 | 1215 | 67 | 0.948 | 1283 | 1223 | 60 | 0.953 | 1253 | | HOC i.p. bend |
| 1185 | 1117 | 68 | 0.943 | 1178 | 1119 | 59 | 0.950 | 1179 | 1123 | 56 | 0.953 | 1159 | 1184 | HOC i.p. bend |
| 992 | 965 | 27 | 0.973 | 994 | 966 | 28 | 0.972 | 996 | 969 | 27 | 0.973 | 960 | | C-O stretch |
| 798 | 787 | 11 | 0.986 | 804 | 795 | 9 | 0.988 | 807 | 798 | 9 | 0.989 | 798 | 784 | $CO_3$ o.p. bend |
| 610 | 601 | 9 | 0.985 | 618 | 610 | 8 | 0.988 | 620 | 614 | 6 | 0.991 | 623 | | OCO bend |
| 582 | 617 | -35 | 1.059 | 575 | 610 | -35 | 1.061 | 576 | 632 | -56 | 1.097 | 550 | | HOC o.p. bend |
| 549 | 527 | 22 | 0.960 | 552 | 536 | 16 | 0.971 | 553 | 540 | 13 | 0.976 | 533 | | OCO bend |
| 514 | 511 | 3 | 0.994 | 506 | 491 | 15 | 0.970 | 507 | 529 | -22 | 1.043 | 458 | | HOC o.p. bend |

[a] Tossell's theoretical predictions from calculations at the CCSD(T)/6-311+G(2d,p) level of theory with corrections for anharmonicity at the B3LYP/CBSB7 level of theory.[12]
[b] Referring to the experimental data obtained by IR spectroscopy of trapped gaseous carbonic acid in a Ne matrix, see the paper of Bernard et al.[9]

In case of the carbonic acid cis-trans monomer the agreement between the ωB97XD/aug-cc-pVQZ calculations and the experiment are rather poor. The deviations are +111, +37, -41, -61 and +14 cm$^{-1}$ for the O-H (~3600 cm$^{-1}$), C=O (~1800 cm$^{-1}$), C-OH stretching (~1400 cm$^{-1}$), C-O-H in-plane bending (~1200 cm$^{-1}$) and $CO^3$ out-of-plane bending (~800 cm$^{-1}$) modes, respectively, where the plus and minus signs indicate whether the experimental frequencies are over- or under-estimated. Tossell's results agree better with the experimental values, although with deviations up to 32 cm$^{-1}$, which appears quite remarkable regarding the high level of theory used. As in the foregoing case there are large variations between the basis sets and only slow convergence with increasing basis set size. The error compensation mentioned above is also present here. In case of the H-O-C out-of-plane bending mode at 490-530 cm$^{-1}$ the results obtained with different basis sets deviate from each other even qualitatively. The double- and triple-zeta basis sets give redshifts of 3 and 15 cm$^{-1}$, respectively, whereas the quadruple-zeta basis yields a blueshift of 22 cm$^{-1}$. The fundamental to harmonic frequency ratios are distributed in the interval 0.94-1.1, but we observe no significant outliers as for the H-O-C out-of-plane bending modes in the cis-cis monomer. Comparing our quadruple-zeta basis set results with Tossell's predictions we note that except for four vibrational modes they agree within at least 40 cm$^{-1}$. Concerning the O-H stretching and H-O-C out-of-plane bending modes they differ by less than 100 cm$^{-1}$.

In case of the dimer, we observe that the fundamental frequencies using the ωB97XD/aug-cc-pVTZ level of theory agree quite well to the experimental data and are slightly better than those obtained by Tossell at the B3LYP/CBSB7 level of theory, except for O-H stretching and C-O-H in-plane bending modes. The deviations from experiment are +57, -30, -49, +3, +19, -92, -86, +6 and +5 cm$^{-1}$ for the O-H stretching, C=O stretching, C-OH stretching, C-O-H in-plane bending and $CO_3$ out-of-plane bending modes, respectively. In case of the O-H stretching mode Tossell achieved results very close to experiment, i.e. he obtained 3624 cm$^{-1}$ deviating only by +1cm$^{-1}$ from experiment. In case of the C-O-H in-plane bending modes, his method performed slightly better than ours too, but still deviates from the experimental result of 1274 cm$^{-1}$ by -70 and -73 cm$^{-1}$ (-92 and -86 cm$^{-1}$ in our case). Concerning those vibrational modes for which no experimental data has been available[9] the differences between our and Tossell's results are small except for the O-H bending and the asymmetric H-bond stretching mode.

The harmonic to anharmonic frequency shifts decrease quite significantly when increasing the basis set size from double- to triple-zeta. We observe qualitative differences too: we count seven vibrational modes for which the results obtained with the aug-cc-pVTZ basis set yield a blueshift, whereas considerable redshifts are obtained using aug-cc-pVDZ. Due to the large anharmonicity of the H-bond stretching, the difference between the two basis sets are most pronounced there. The double-zeta basis yields frequency redshifts up to 636 and 726 cm$^{-1}$ and fundamental frequencies of 2631 and 2434 cm$^{-1}$ for symmetric and asymmetric H-bond vibrations, respectively, whereas the redshifts obtained with the triple-zeta basis are 447 and 525 cm$^{-1}$ leading to fundamental frequencies of 2801 and 2617 cm$^{-1}$.

In section 2 we described variational calculations of these H-bond vibrations by means of the discrete variable representation (DVR) method. These frequencies are not as basis set dependent as those obtained with the perturbative estimation of fundamental frequencies above. The frequency shifts are 395 and 441 (D), 386 and 442 (T) and 384 and 428 (Q) cm$^{-1}$ for the symmetric and asymmetric H-bond vibrations in potential energy surfaces calculated at the ωB97XD/aug-cc-pVXZ, X=D, T, Q, levels, respectively. The results obtained with the perturbative treatment at the ωB97XD/aug-cc-pVTZ level of theory are much closer to these values of frequency shifts than the results obtained with the smaller double-zeta basis set. We conclude that also here the double-zeta basis is much too small to yield converged results for the fundamental frequencies. Although considerably better, the results from the triple-zeta basis are thought to be still rather far from being converged and present rather a semiquantitative estimate of the fundamental frequencies than an accurate theoretical prediction. This can be seen as well from the deviations between experimental values and Tossell's theoretical results. Tossell's predictions of the fundamental H-bond frequencies with the B3LYP/CBSB7 method and a perturbative treatment for anharmonicity are close to our variational DVR results using the harmonic frequencies at the ωB97XD/aug-cc-pVXZ, X=D, T, level of theory, see TABLE IV. Our results deviate from his predictions by -11 and -23 cm$^{-1}$ for symmetric and asymmetric H-bond stretching modes in case of the double-zeta basis, and by -21 and -42 cm$^{-1}$ in case of the triple-zeta basis. As in the case of the cis-trans monomer we observe that especially for hydrogen stretching modes B3LYP appears to be well suited to estimate fundamental frequencies, whereas the ωB97XD functional seems to converge from below to slightly overestimated fundamental frequencies.

**TABLE IV.** Values like in TABLE III for the carbonic acid dimer. The values obtained with the DVR method are given in parenthesis.

| ωB97XD/aug-cc-pVDZ | | | | ωB97XD/aug-cc-pVTZ | | | | Tossell[a] | Expt.[b] | Vib. mode |
|---|---|---|---|---|---|---|---|---|---|---|
| Harm. | Fund. | Shift | Ratio | Harm. | Fund. | Shift | Ratio | | | |
| 3868 | 3663 | 205 | 0.947 | 3865 | 3680 | 185 | 0.952 | 3624 | 3623 (2500-3500) | O-H stretch |
| 3868 | 3662 | 204 | 0.947 | 3864 | 3680 | 184 | 0.952 | 3624 | 3623 | O-H stretch |
| 3267 | 2631 (2872) | 636 (395) | 0.805 | 3248 | 2801 (2862) | 447 (386) | 0.862 | 2883 | | H-bond stretch |
| 3160 | 2434 (2729) | 726 (441) | 0.770 | 3142 | 2617 (2710) | 525 (432) | 0.833 | 2752 | | H-bond stretch |
| 1818 | 1764 | 54 | 0.971 | 1816 | 1769 | 47 | 0.975 | 1752 | 1799 (1700) | C=O stretch |
| 1747 | 1658 | 89 | 0.949 | 1742 | 1678 | 64 | 0.963 | 1656 | 1727 (1608) | C=O stretch |
| 1586 | 1548 | 38 | 0.976 | 1585 | 1568 | 17 | 0.990 | 1519 | 1565 (1530) | C-OH stretch |
| 1551 | 1517 | 34 | 0.978 | 1550 | 1534 | 16 | 0.990 | 1478 | 1515 (1502) | C-OH stretch |
| 1411 | 1402 | 9 | 0.994 | 1409 | 1419 | -10 | 1.008 | 1386 | (1405) | HOC i.p. bend |
| 1407 | 1397 | 10 | 0.993 | 1402 | 1414 | -12 | 1.009 | 1343 | (1297) | HOC i.p. bend |
| 1260 | 1183 | 77 | 0.939 | 1250 | 1182 | 68 | 0.946 | 1204 | 1274 | HOC i.p. bend |
| 1256 | 1184 | 72 | 0.942 | 1246 | 1188 | 58 | 0.953 | 1201 | 1274 | HOC i.p. bend |
| 1052 | 1027 | 25 | 0.976 | 1056 | 1046 | 10 | 0.990 | 1013 | (1054) | C-O stretch |
| 1044 | 1029 | 15 | 0.986 | 1048 | 1040 | 8 | 0.992 | 1006 | (1035) | C-O stretch |
| 1000 | 968 | 32 | 0.968 | 999 | 964 | 35 | 0.966 | 947 | | O-H---O o.p. bend |
| 959 | 922 | 37 | 0.962 | 957 | 923 | 34 | 0.964 | 897 | (884) | O-H---O o.p. bend |
| 820 | 808 | 12 | 0.985 | 825 | 817 | 8 | 0.990 | 795 | 811 (812) | CO$_3$ o.p. bend |
| 819 | 807 | 12 | 0.985 | 825 | 816 | 9 | 0.989 | 794 | 811 (812) | CO$_3$ o.p. bend |
| 657 | 663 | -6 | 1.008 | 663 | 676 | -14 | 1.021 | 648 | (683) | O=C-O i.p. bend |
| 647 | 613 | 34 | 0.948 | 652 | 666 | -14 | 1.021 | 639 | (658) | O=C-O i.p. bend |
| 611 | 539 | 72 | 0.882 | 613 | 625 | -12 | 1.019 | 598 | (657) | O=C-O i.p. bend |
| 597 | 324 | 273 | 0.543 | 588 | 448 | 140 | 0.762 | 574 | | HOC o.p. bend |
| 595 | 324 | 271 | 0.544 | 586 | 448 | 138 | 0.763 | 573 | | HOC o.p. bend |
| 576 | 559 | 17 | 0.971 | 581 | 570 | 11 | 0.982 | 566 | | O=C-O i.p. bend |
| 214 | 108 | 106 | 0.505 | 209 | 256 | -47 | 1.228 | 209 | | Interm. i.p. shear |
| 201 | 35 | 166 | 0.176 | 193 | 222 | -29 | 1.147 | 185 | | Interm. stretch |
| 171 | 156 | 15 | 0.914 | 169 | 185 | -16 | 1.095 | 161 | | Interm. i.p. shear |
| 128 | 119 | 9 | 0.928 | 124 | 120 | 4 | 0.964 | 121 | | Interm. o.p. shear |
| 76 | 46 | 30 | 0.605 | 74 | 74 | 0 | 0.998 | 76 | | Interm. o.p. shear |
| 59 | 46 | 13 | 0.783 | 56 | 52 | 4 | 0.945 | 60 | | Interm. o.p. shear |

[a] Tossell's predictions from calculations at the CCSD(T)/6-311+G(2d,p) level of theory with corrections for anharmonicity at the B3LYP/CBSB7 level of theory.[12]
[b] Referring to the experimental data obtained by IR spectroscopy of trapped gaseous carbonic acid in a Ne matrix, see the paper of Bernard et al.[9] For comparison, the values for solid cryogenic ß-H$_2$CO$_3$, which has a one dimensional H-bonded chain structure, are given in parenthesis and have been adopted from the paper of Kohl et al.[37] In the latter work, the authors report without assignment spectral lines at 605, 307, 258 and 193 cm$^{-1}$.

**C. Stability and energetics of the dimer**

We discuss the energetics of the formation reaction of the carbonic acid dimer

$$2H_2O + 2CO_2 \rightarrow (H_2CO_3)_2$$

at different levels of theory. Results for reaction energies (reaction enthalpies at zero temperature), are summarized in TABLE V. They include also counterpoise-corrected values for most cases and the zero-point energy $\Delta$ZPE for those levels where at least harmonic frequencies could be calculated. In a first step a reference value for the reaction energy was calculated at the G4(MP2) level of theory[28] yielding 0.37 kcal/mol. Even if the average accuracy of about 1 kcal/mol assessed on the G3/05 test set[28] with this method is assumed as granted for individual cases, we note that it is not possible to draw a definite conclusion on whether the reaction energy is slightly negative or positive. Results from combinations of the aug-cc-pVXZ, X=D, T, Q basis sets and the B3LYP, MP2, ωB97XD density functionals are not conclusive too. B3LYP yields slightly negative results with the double-zeta basis (-1.99 kcal/mol), but slightly positive ones with the larger basis sets (1.73 and 2.22 kcal/mol), where the energies tend to increase with increasing basis set size and thus appear to converge to a positive value in the limit of an infinite basis set. Taking into account counterpoise corrections using the quadruple-zeta basis destabilizes the binding only by 0.14 kcal/mol. The contrary is true for the MP2 method. The energy obtained with the double-zeta basis set is quite significantly positive with 5.59 kcal/mol, but is significantly reduced by going to the triple-zeta basis to 1.9 kcal/mol. Note that for estimating the zero-point energy in case of the MP2/aug-cc-pVTZ level of theory, the value of $\Delta$ZPE at the MP2/aug-cc-pVDZ level of theory has been taken based on the observation that $\Delta$ZPE does only depend very weakly on basis set size.[11,13] In this case, the counterpoise correction yields a significantly different result of 3.93 kcal/mol. The reason for this is probably again that for MP2 the triple-zeta basis is still too small to yield converged results. It has even been argued that one should not rely too heavily on counterpoise corrections when estimating the formation energy of the carbonic acid dimer.[11] The ωB97XD functional follows the trend of the B3LYP functional, i.e. the energies increase with increasing basis set size. However, all three energies obtained with ωB97XD, i.e. -9.65, -6.22 and -5.48 kcal/mol for the double-, triple- and quadruple-zeta basis, respectively, are clearly negative and seem to converge. The counterpoise correction using the quadruple-zeta basis gives rise to a just slightly less exothermic reaction energy of -5.07 kcal/mol as in case of B3LYP. In order to reach a definite conclusion on the formation energy of the carbonic acid dimer we further investigated the energetics with a combination of methods that could be denoted as CCSD(T)/aug-cc-pVYZ//METHOD/aug-cc-pVXZ. The geometry of the molecule was optimized and harmonic frequencies were calculated at the METHOD/aug-cc-pVXZ level of theory, where METHOD=B3LYP, MP2, ωB97XD and X=Q for B3LYP and ωB97XD and X=T for MP2 (harmonic frequency calculations used the aug-cc-pVDZ basis set, see above), and subsequently a single-point calculation was performed at the optimized geometry at the CCSD(T)/aug-cc-pVYZ level of theory, where Y=D, T. The results with the double-zeta basis set (CCSD(T)/aug-cc-pVDZ//METHOD/aug-cc-pVXZ) are still inclusive. Although all energies with respect to different cis/trans geometries are slightly negative (-1.57, -0.07, -1.21 kcal/mol), the counterpoise corrections are quite large and lead to positive values (1.61, 3.18, 1.97 kcal/mol). The reason for the large effect of the counterpoise corrections is thought to be the same as with the MP2 method above, i.e. the insufficient size of the basis. However, at the CCSD(T)/aug-cc-pVTZ//METHOD/aug-cc-pVXZ level of theory both counterpoise-corrected and non-counterpoise-corrected reaction energies are clearly negative. Extrapolating these results (using Truhlar's extrapolation scheme[30]) to the CCSD(T)/aug-cc-pVQZ//METHOD/aug-cc-pVXZ level of theory and including Hartree-Fock corrections, see section 2, we arrive at our best estimates for the reaction energy of -5.72, -5.57, -5.60 kcal/mol without counterpoise corrections and -4.80, -4.64, -4.68 kcal/mol with counterpoise corrections for the B3LYP, MP2 and ωB97XD methods, respectively. Interestingly, these values are close to the reaction energies predicted at the ωB97XD/aug-cc-pVQZ level of theory at considerably lower computational cost. Including zero-point energies due to anharmonic vibrations in case of the ωB97XD as calculated with the aug-cc-pVTZ basis set, see section 3.2, we arrive at -6.05 kcal/mol without the counterpoise correction and -5.14 kcal/mol with the counterpoise correction. These values bracket our best estimate for the energy of the reaction which we believe now to be clearly exothermic.

The accuracy of our best estimate can be considered as a significant improvement over, for example, the G4(MP2) extrapolation scheme for the following reasons:

The optimization and frequency calculations use the aug-cc-pVQZ basis set, which yields 824 basis functions in case of the dimer, in contrast to the 6-31G(2df,p) basis set,[38] which yields 244 basis functions in case of the dimer. In addition, the ωB97XD functional is used, which appears to be especially suited for molecules containing weak H-bonds due to the inclusion of long-range corrections and empirical dispersion. Moreover, fundamental frequencies are estimated to incorporate anharmonic corrections into the zero-point energy instead of using the harmonic approximation.

**TABLE V.** Reaction energies with and without counterpoise corrections, and differences in zero-point energies, ΔZPE, for the reaction $2H_2O + 2CO_2 \rightarrow (H_2CO_3)_2$ at different levels of theory. All energies are given kcal/mol.

| Level of Theory | Reaction energy: $2H_2O + 2CO_2 \rightarrow (H_2CO_3)_2$ | Counterpoise corrected reaction energy | ΔZPE |
|---|---|---|---|
| G4(MP2) | 0.37 | | 9.30 |
| HF//B3LYP/aug-cc-pVQZ | 3.92 | 4.02 | |
| HF/aug-ccpV5Z//B3LYP/aug-cc-pVQZ | 4.07 | 4.08 | |
| HF/aug-ccpV6Z//B3LYP/aug-cc-pVQZ (extrap.) | 4.10 | 4.09 | |
| HF//MP2/aug-cc-pVTZ | 1.65 | 1.75 | |
| HF/aug-ccpV5Z//MP2/aug-cc-pVTZ | 1.78 | 1.79 | |
| HF/aug-ccpV6Z//MP2/aug-cc-pVTZ (extrap.) | 1.81 | 1.90 | |
| HF//wB97XD/aug-cc-pVQZ | 3.91 | 4.02 | |
| HF/aug-ccpV5Z//wB97XD/aug-cc-pVQZ | 4.06 | 4.07 | |
| HF/aug-ccpV6Z//wB97XD/aug-cc-pVQZ (extrap.) | 4.10 | 4.08 | |
| B3LYP/aug-cc-pVDZ | -1.99 | | 9.72 |
| B3LYP/aug-cc-pVTZ | 1.73 | | 9.57 |
| B3LYP/aug-cc-pVQZ | 2.22 | 2.36 | 9.56 |
| MP2/aug-cc-pVDZ | 5.59 | | 9.99 |
| MP2/aug-cc-pVTZ | 1.90 | 3.93 | Not calc. |
| wB97XD/aug-cc-pVDZ | -9.65 | | 10.10 (8.68) |
| wB97XD/aug-cc-pVTZ | -6.22 | | 9.91 (9.49) |
| wB97XD/aug-cc-pVQZ | -5.48 | -5.07 | 9.94 |
| CCSD(T)/aug-cc-pVDZ//B3LYP/aug-cc-pVQZ | -1.57 | 1.61 | |
| CCSD(T)/aug-cc-pVDZ//MP2/aug-cc-pVTZ | -0.07 | 3.18 | |
| CCSD(T)/aug-cc-pVDZ//wB97XD/aug-cc-pVQZ | -1.21 | 1.97 | |
| CCSD(T)/aug-cc-pVTZ//B3LYP/aug-cc-pVQZ | -4.48 | -2.72 | |
| CCSD(T)/aug-cc-pVTZ//MP2/aug-cc-pVTZ | -3.92 | -2.13 | |
| CCSD(T)/aug-cc-pVTZ//wB97XD/aug-cc-pVQZ | -4.28 | -2.52 | |
| CCSD(T)/aug-cc-pVQZ//B3LYP/aug-cc-pVQZ (extrap.) | -6.41 | -5.49 | |
| CCSD(T)/aug-cc-pVQZ//MP2/aug-cc-pVTZ (extrap.) | -6.29 | -5.34 | |
| CCSD(T)/aug-cc-pVQZ//wB97XD/aug-cc-pVQZ (extrap.) | -6.30 | -5.37 | |
| CCSD(T)/aug-cc-pVQZ//B3LYP/aug-cc-pVQZ (extrap.) + dHF | -5.72 | -4.80 | |
| CCSD(T)/aug-cc-pVQZ//MP2/aug-cc-pVTZ (extrap.) + dHF | -5.57 | -4.64 | |
| CCSD(T)/aug-cc-pVQZ//wB97XD/aug-cc-pVQZ (extrap.) + dHF | -5.60 | -4.68 | |
| CCSD(T)/aug-cc-pVQZ//wB97XD/aug-cc-pVQZ (extrap.) + dHF + dZPE | -6.05 | -5.14 | |

The basis energy is calculated at the CCSD(T)/aug-cc-pVXZ, X=D,T (yielding 220 and 460 basis functions in case of the dimer), levels of theory and extrapolated to the CCSD(T)/aug-cc-pVQZ level of theory, whereas using G4(MP2) the basis energy is calculated only at the CCSD(T)/6-31G(d) level of theory (yielding only 80 basis functions in case of the dimer) using in addition the frozen core option in the calculation. It is not expected that the MP2 correction at the MP2/G3LargeXP level of theory, which yields 188 basis functions in case of the dimer can overcome these differences in basis set sizes.

The HF corrections use the aug-cc-pVQZ (824 basis functions in case of the dimer) and the aug-cc-pV5Z (1336 basis functions in case of the dimer) basis sets instead of the aug-cc-pVTZ (460 basis functions in case of the dimer) and the aug-cc-pVQZ basis sets as used in G4(MP2) theory to calculate HF corrections.

Higher level corrections[28] cancel as the number of beta electrons is the same in the products and reactants of the dimer formation reaction from water and carbon dioxide. Spin-orbit corrections[27] are included in G4(MP2) only for atoms and diatomic molecules, and are thus making no difference for this reaction.

We conclude this section by reporting on our best estimates for the formation energies of the two monomers under consideration, which are 6.08 kcal/mol in case of the cis-cis monomer and 7.65 kcal/mol in case of the cis-trans monomer. These results can be regarded as slight improvements on the values reported very recently.[14]

It should be noted that the relevant quantity if discussing equilibria is, of course, the free energy of reaction, which is only partly given by the reaction energies discussed above. It should be kept in mind that the free energy of dimerization is more positive than the reaction energy due to the reduction of two molecules to one and is expected to be strongly temperature dependent.

**D. Detection of Carbonic Acid**

Water and carbon dioxide ices exist in various different environments in outer space. Carbonic acid is expected to be present in these environments as well.[5,6,8,9,14,39] It has been suggested that solid carbonic acid may be found on the Martian surface, on interstellar grains or on Jupiter's satellites Europa, Ganymede, and Callisto. Gaseous carbonic acid is expected to exist in our own atmosphere,[5] in the atmosphere of Mars and Venus and even on comets.[5,9] So far, however, no spectroscopic detections of carbonic acid have been reported in any of these environments.

In the following we discuss the reasons for this and the possibilities to detect carbonic acid via spectroscopic observations in the gas phase in the two most promising environments, namely Earth's atmosphere and Mars.

In Earth's atmosphere it is thought that carbonic acid may be formed in the upper troposphere, most likely in cirrus clouds.[5] Cirrus clouds typically occur in heights between 8 and 12 km, where one finds temperatures around 250 K and a pressure around 250 mbar. Following Hage et al.,[5] carbonic acid can be sublimated and recondensed without decomposition into $CO_2$ and $H_2O$. The temperature and pressure range relevant for the sublimation in the laboratory has been 180 to 220 K and $10^{-7}$ to $10^{-5}$ mbar. In our atmosphere we find regions with this temperature in a height of about 15 to 20 km, which is already above the highest cirrus clouds. Looking for a region with a pressure around $10^{-5}$ mbar, we have already to go up to heights over 120 km. Although we expect carbonic acid to be found even at higher temperatures and pressure, like in the cirrus clouds, the collision rate is expected to be much higher leading to a much lower abundance of carbonic acid. Thus, it might be very challenging to detect carbonic acid in our atmosphere. Additionally, the detection is complicated by the fact that the spectroscopic lines of carbonic acid are not in a preferable part of the spectrum. We compared our model spectra with the atmospheric spectrum including all known molecules with regard to the relevant conditions, see FIG. 2. This sky model spectrum was calculated using the Line-By-Line-Radiative-Transfer-Model (LBLRTM V12.0)[40] code package, incorporating HITRAN 2008[41] as line database. The sky model is calculated for Cerro Paranal (24° 37' 33" S, 70° 24' 11" W, 2635m, with an airmass=1.0), the location of the Very Large Telescope of the European Southern Observatory (ESO)[42] in the Chilenian Atacama desert. We used an equatorial standard atmospheric profile[43] containing height profiles for a large number of molecules, which was created by using observational data obtained by the Michelson Interferometer for Passive Atmospheric Sounding (MIPAS) instrument on-board the Envisat satellite. The intensity was normalized by an exponential factor representing the Rayleigh-Taylor approximation of the thermal emission of the atmosphere. In the central region of the graph the normalizing factor is about $2.7 \times 10^6$ photons per second per $m^2$ per µm per $arcsec^2$. As the Atacama desert is known as one of the driest places we refined this equatorial profile by additionally selecting appropriate data from the Global Data Assimilation System (GDAS[44] provided by the National Oceanic and Atmospheric Administration, NOAA[45]). These meteorological data are available on a 3 hour basis, as a global, 1 degree latitude/longitude data set (from December 2004 to present and ongoing), and contain height profiles for the temperature, pressure and relative humidity. In addition, the on-site meteo monitor of ESO was also incorporated to finally create an averaged atmospheric profile appropriate for Cerro Paranal. Unfortunately the spectral lines of carbonic acid are in a range where gluts of lines originating from other molecules are certain to be found. We were especially searching for the spectral feature of the O-H stretch of the carbonic acid dimer, which is expected to be the most prominent one. There are a few windows in the atmospheric spectrum, where the atmosphere is extensively transparent. There is no intermediate absorption and re-emission. In the range from about 4000 $cm^{-1}$ to higher wavenumbers there is the astronomical K band (FIG. 2, right part). In the K band there is hardly any continuum emission, therefore the background is very small. The range from about 3500 cm$^{-1}$ to lower wavenumbers is the L band, which has already more continuum emission compared to the K band, but at least there are a few emission-free gaps. As one can see in FIG. 2, the spectral line of the most important O-H stretching mode of the carbonic acid dimer is just in between the K and the L band. In this range the largest contributions to the atmospheric spectrum according to our calculations originate from $H_2O$, $CH_4$, and $CO_2$. We investigated all other spectral features of carbonic acid dimer and both monomers as well. They are all at lower wavenumbers, even further in the thermal infrared, where the background radiation increases dramatically by two orders of magnitudes. We have to conclude that we have hardly any chance to detect carbonic acid in our atmosphere with present spectroscopic methods and instruments.

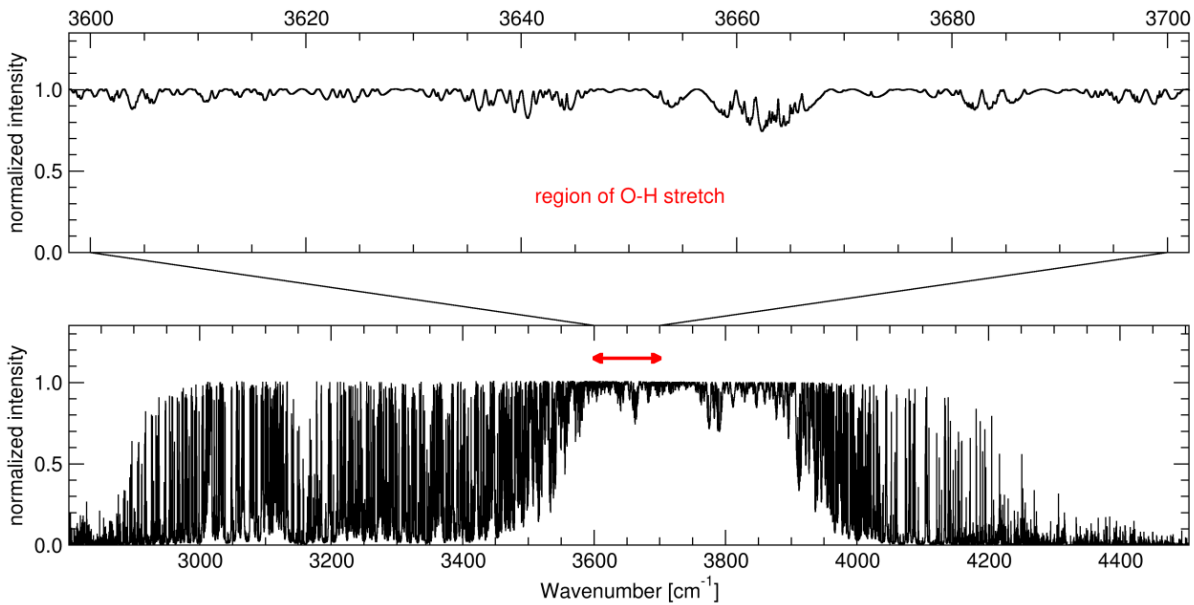

FIG. 2. The sky model spectrum calculated using the LBLRTM V12.0[40] in the relevant range of the expected strongest line of the carbonic acid dimer.

On Mars one can find carbon dioxide ice and water ice at both polar caps. Wu et al. suggest that this mixture could be a source for the production of carbonic acid.[14] Since Mars has no global magnetosphere any more, high energy solar photons and cosmic rays could provide the energy required to form solid carbonic acid. Through sublimation and recondensation, these molecules could be released into the atmosphere. During the pole's summer, the carbon dioxide and water sublimes into the atmosphere. In these seasons, large amounts of dust and water vapor are located in the Martian atmosphere. Large cirrus clouds are formed consisting of both water ice and carbon dioxide. The water ice clouds are in a height of about 20 km, while the carbon dioxide ice clouds are in higher regions of about 50 km. The pressure in these heights is below 1 mbar at 20 km and even lowers to about $10^{-3}$ mbar at 60 km. The temperature range is between 100 and 200 K depending on height and solar irradiation. Thus, the conditions for the detection of carbonic acid are promising.

In the following, we discuss the possibility of observations from ground based telescopes or satellites. Direct measurements of the atmosphere with the Martian surface in the background seem impossible because of the self-radiation of the planet. The spectroscopic measurements would be completely overlaid. For this reason, the possibility of grazing observations was taken into account. The thickness of the atmosphere that has to be observed is about 50 km. Taking the smallest possible distance Mars – Earth, which is about $56 \times 10^6$ km, leads to a required resolution of at least 0.1 arcsec. E.g., the Hubble Space Telescope (HST) has this resolution in the optical part, the resolution in the required infrared is worse. With all present instruments it is not possible to achieve the required resolution. The next generation space telescope James Webb Space Telescope (JWST) will be in the same range as HST, so it seems we have to wait for the E-ELT, the European Extremely Large Telescope. With this telescope we will reach a resolution below 1 milliarcsec. First light for the E-ELT is expected in 2022.[46]

Another option is space exploration missions. The planned mission MAVEN (Mars Atmosphere and Volatile EvolutioN) will be launched in late 2013.[47] The aim of the mission is collecting data about the Martian atmosphere. During this mission a measurement and detection of carbonic acid seems to be feasible.

In order to search for carbonic acid in gas phase in other astrophysical environments like the interstellar matter (ISM) or Jupiter's satellites, further studies should concentrate on providing high-resolution infrared spectra from laboratory measurements and investigations on temperature and pressure effects. These data are essential for extended future spectroscopic investigations in outer space.

## IV. Conclusions

We find that, due to cancellation of errors, the B3LYP and MP2 methods lead to harmonic frequencies which are close to experimentally determined fundamental ones.. A more systematic prediction of fundamental frequencies based on the ωB97XD functional led to good agreement with experimental results for the carbonic acid cis-cis monomer, but only to a rough estimate for the cis-trans monomer and the carbonic acid dimer under consideration. Nevertheless we observe a slightly better agreement with experiment than in earlier theoretical predictions where the B3LYP functional was employed for most vibrational modes of the carbonic acid dimer. B3LYP appears to perform very well for O-H stretching modes in the region around 3600 cm$^{-1}$ and H-bond vibrations at 2600-2800 cm$^{-1}$. For this reason, both methods seem to provide a qualitatively good first estimate of fundamental frequencies. Basis sets of triple-zeta quality or larger are necessary in any case. Obviously, the estimation of accurate fundamental frequencies remains a difficult task and if aiming for reliable quantitative predictions, one has to go to significantly higher levels of theory. This has been shown in the comparison of theoretical and experimental data in case of the cis-trans monomer. Though the results of Tossell[12] were obtained at a high level of theory and reproduce the experimental data quite nicely, even they still exhibit differences of up to 40 cm$^{-1}$.

A re-investigation of the energetics of the formation of the carbonic acid dimer from water and carbon dioxide has affirmed that carbonic acid is not only experimentally a very elusive molecule. It turned out that a quite high level of theory is needed to draw a definite conclusion on the 15 years old question[11] whether the carbonic acid dimer is more stable than its constituents water and carbon dioxide. At least according to our best estimate this question can be answered with yes. Furthermore, we note that the clearly negative reaction enthalpy of about -5.14 kcal/mol at zero temperature can be obtained quite accurately at the ωB97XD/aug-cc-pVQZ level of theory (leading to -5.48 kcal/mol) at much reduced computational cost compared to CCSD(T) and related methods. This confirms the suitability of that functional, especially for molecules containing weak bonds such as hydrogen bonds.

Despite its presence is very likely, the detection of carbonic acid in astrophysical environments seems quite challenging. With present spectroscopic methods and instruments it will not be possible to detect carbonic acid in Earth's atmosphere. An upcoming space exploration mission makes it imaginable to detect it in the Martian atmosphere.

**Acknowledgement**

This work was supported by the Austrian Ministry of Science BMWF as part of the UniInfrastrukturprogramm of the Research Platform Scientific Computing at the University of Innsbruck. S.H. and S.D. are funded by Austrian Science Fund (FWF) DK+ project Computational Interdisciplinary Modeling, W1227-N16. W.K. is funded by Austrian ESO In-Kind project funded by BM:wf under contracts BMWF-10.490/0009-II/10/2009 and BMWF-10.490/0008-II/3/2011.

**References**

1. Z. S. Fisher, M. C. Maupin, M. Budayova-Spano, L. Govindasamy, C. Tu, M. Agbandje-McKenna, D. N. Silverman, G. A. Voth, and R. McKenna, Biochemistry **46** (11), 2930 (2007); S. Thoms, Journal of Theoretical Biology **215** (4), 399 (2002).
2. D. M. Kern, Journal of Chemical Education **37** (1), 14 (1960).
3. R. A. Feely, C. L. Sabine, K. Lee, W. Berelson, J. Kleypas, V. J. Fabry, and F. J. Millero, Science **305** (5682), 362 (2004); C. L. Sabine, R. A. Feely, N. Gruber, R. M. Key, K. Lee, J. L. Bullister, R. Wanninkhof, C. S. Wong, D. W. R. Wallace, B. Tilbrook, and e. al, Science **305** (5682), 367 (2004); J. C. Orr, V. J. Fabry, O. Aumont, L. Bopp, S. C. Doney, R. A. Feely, A. Gnanadesikan, N. Gruber, A. Ishida, F. Joos, and e. al, Nature **437** (7059), 681 (2005).
4. H. A. Al-Hosney and V. H. Grassian, Journal of the American Chemical Society **126** (26), 8068 (2004); H. A. Al-Hosney and V. H. Grassian, Physical Chemistry Chemical Physics **7** (6), 1266 (2005); H. A. Al-Hosney, S. Carlos-Cuellar, J. Baltrusaitis, and V. H. Grassian, Physical Chemistry Chemical Physics **7** (20), 3587 (2005).
5. W. Hage, K. R. Liedl, A. Hallbrucker, and E. Mayer, Science **279**, 1332 (1998).
6. W. Zheng and R. I. Kaiser, Chemical Physics Letters **450** (1-3), 55 (2007).
7. J. K. Terlouw and H. Schwarz, Angewandte Chemie **99** (9), 829 (1987).
8. T. Mori, K. Suma, Y. Sumiyoshi, and Y. Endo, The Journal of Chemical Physics **130**, 204308/1 (2009).
9. J. Bernard, M. Seidl, I. Kohl, K. R. Liedl, E. Mayer, O. Galvez, H. Grothe, and T. Loerting, Angewandte Chemie International Edition **50**, 1939 (2011).
10. C. A. Wight and A. I. Boldyrev, Journal of Chemical Physics **99**, 12125 (1995).
11. K. R. Liedl, S. Sekusak, and E. Mayer, Journal of the American Chemical Society **119**, 3782 (1997).
12. J. A. Tossell, Inorganic Chemistry **45**, 5961 (2006).
13. J. Murillo, J. David, and A. Restrepo, Physical Chemistry Chemical Physics **12**, 10963 (2010).
14. Y.-J. Wu, R. C. Y. Wu, and M.-C. Liang, Icarus **214**, 228 (2011).
15. A. Becke, Journal of Chemical Physics **98** (7), 5648 (1993).
16. C. Lee, W. Yang, and R. G. Parr, Physical Review B: Condensed Matter and Materials Physics **37**, 785 (1988).
17. J. Chai and M. Head-Gordon, Physical Chemistry Chemical Physics **10**, 6615 (2008).
18. C. Moller and M. S. Plesset, Physical Review **46**, 618 (1934); M. Head-Gordon, J. A. Pople, and M. J. Frisch, Chemical Physics Letters **153**, 503 (1988).
19. T. H. Dunning Jr., Journal of Chemical Physics **90**, 1007 (1989); R. A. Kendall, T. H. Dunning Jr., and J. R. Harrison, Journal of Chemical Physics **96**, 6796 (1992); K. A. Peterson, D. E. Woon, and T. H. Dunning Jr., Journal of Chemical Physics **100**, 7410 (1994); E. R. Davidson, Chemical Physics Letters **260**, 514 (1996).
20. W. H. Miller, N. C. Handy, and J. E. Adams, Journal of Chemical Physics **72** (1), 99 (1980).
21. V. Barone and C. Minichino, Journal of Molecular Structure **330**, 365 (1995).
22. V. Barone, The Journal of Chemical Physics **122**, 014108/1 (2005).
23. K. Hermansson, M. M. Probst, G. Gajewski, and P. D. Mitev, The Journal of Chemical Physics **131**, 244517/1 (2009).
24. J. C. Light, I. P. Hamilton, and J. V. Lill, The Journal of Chemical Physics **82** (3), 1400 (1985); R. Dawes and T. Carrington Jr., Journal of Chemical Physics **121** (2), 726 (2004).
25. I. Wolfram Research, *Mathematica Edition: Version 8.0*. (Wolfram Research, Inc., Champaign, Illinois, 2010).
26. J. A. Pople, M. Head-Gordon, D. J. Fox, K. Raghavachari, and L. A. Curtiss, Journal of Chemical Physics **90**, 5622 (1989); L. A. Curtiss, K. Raghavachari, G. W. Trucks, and J. A. Pople, Journal of Chemical Physics **94**, 7221 (1991); L. A. Curtiss, K. Raghavachari, P. C. Redfern, V. Rassolov, and J. A. Pople, Journal of Chemical Physics **109**, 7764 (1998).
27. L. A. Curtiss, P. C. Redfern, and K. Raghavachari, The Journal of Chemical Physics **126**, 084108/1 (2007).
28. L. A. Curtiss, P. C. Redfern, and K. Raghavachari, The Journal of Chemical Physics **127**, 124105/1 (2007).
29. J. A. Pople, M. Head-Gordon, and K. Raghavachari, The Journal of Chemical Physics **87**, 5968 (1987).
30. D. G. Truhlar, Chemical Physics Letters **294** (1-3), 45 (1998).
31. A. Halkier, T. Helgaker, P. Jorgensen, W. Klopper, and J. Olsen, Chemical Physics Letters **302**, 437 (1999).
32. M. J. Frisch, G. W. Trucks, H. B. Schlegel, G. E. Scuseria, M. A. Robb, J. R. Cheeseman, G. Scalmani, V. Barone, B. Mennucci, G. A. Petersson, H. Nakatsuji, M. Caricato, X. Li, H. P. Hratchian, A. F. Izmaylov, J.

Bloino, G. Zheng, J. L. Sonnenberg, M. Hada, M. Ehara, K. Toyota, R. Fukuda, J. Hasegawa, M. Ishida, T. Nakajima, Y. Honda, O. Kitao, H. Nakai, T. Vreven, J. A. Montgomery, Jr., J. E. Peralta, F. Ogliaro, M. Bearpark, J. J. Heyd, E. Brothers, K. N. Kudin, V. N. Staroverov, R. Kobayashi, J. Normand, K. Raghavachari, A. Rendell, J. C. Burant, S. S. Iyengar, J. Tomasi, M. Cossi, N. Rega, J. M. Millam, M. Klene, J. E. Knox, J. B. Cross, V. Bakken, C. Adamo, J. Jaramillo, R. Gomperts, R. E. Stratmann, O. Yazyev, A. J. Austin, R. Cammi, C. Pomelli, J. W. Ochterski, R. L. Martin, K. Morokuma, V. G. Zakrzewski, G. A. Voth, P. Salvador, J. J. Dannenberg, S. Dapprich, A. D. Daniels, O. Farkas, J. B. Foresman, J. V. Ortiz, J. Cioslowski, and D. J. Fox, Gaussian 09, Revision A.1 (Gaussian, Inc., Wallingford CT, 2009).

[33] I. Matanovic and N. Doslic, Chemical Physics **338**, 121 (2007); I. Yavuz and C. Trindle, Journal of Chemical Theory and Computation **4** (3), 533 (2008).

[34] J. A. Montgomery, M. J. Frisch, J. W. Ochterski, and G. A. Petersson, The Journal of Chemical Physics **110**, 2822 (1999).

[35] P. Carbonniere, T. Lucca, C. Pouchan, N. Rega, and V. Barone, Journal of Computational Chemistry **26** (4), 384 (2005).

[36] NIST Computational Chemistry Comparison and Benchmark Database, NIST Standard Reference Database 55 Number 101, Release 15b, August 2011, Editor: Russell D. Johnson III, http://cccbdb.nist.gov/

[37] I. Kohl, K. Winkel, M. Bauer, K. R. Liedl, T. Loerting, and E. Mayer, Angewandte Chemie Int. Ed. **48**, 2690 (2009).

[38] R. Ditchfield, W. J. Hehre, and J. A. Pople, The Journal of Chemical Physics **54**, 724 (1971).

[39] T. Loerting and J. Bernard, CHEMPHYSCHEM **11**, 2305 (2010).

[40] S. A. Clough, M. W. Shephard, E. J. Mlawer, J. S. Delamere, M. J. Iacono, K. Cady-Pereira, S. Boukabara, and P. D. Brown, Journal of Quantitative Spectroscopy and Radiative Transfer **91**, 233 (2005); see http://rtweb.aer.com/lblrtm_frame.html

[41] L. S. Rothman, I. E. Gordon, A. Barbe, D. C. Benner, P. F. Bernath, M. Birk, V. Boudon, L. R. Brown, A. Campargue, J.-P. Champion, K. Chance, L. H. Coudert, V. Dana, V. M. Devi, S. Fally, J.-M. Flaud, R. R. Gamache, A. Goldman, D. Jacquemart, I. Kleiner, N. Lacome, W. J. Lafferty, J.-Y. Mandin, S. T. Massie, S. N. Mikhailenko, C. E. Miller, N. Moazzen-Ahmadi, O. V. Naumenko, A. V. Nikitin, J. Orphal, V. I. Perevalov, A. Perrin, A. Predoi-Cross, C. P. Rinsland, M. Rotger, M. Šime?ková, M. A. H. Smith, K. Sung, S. A. Tashkun, J. Tennyson, R. A. Toth, A. C. Vandaele, and J. Vander Auwera, Journal of Quantitative Spectroscopy and Radiative Transfer **110**, 533 (2009); see http://www.cfa.harvard.edu/hitran/

[42] see http://www.eso.org

[43] see http://www.atm.ox.ac.uk/RFM/atm/

[44] see ftp://arlftp.arlhq.noaa.gov/pub/archives/gdas1/

[45] see http://www.noaa.gov/

[46] R. Gilmozzi and M. Kissler-Patig, The Messenger **143**, 25 (2011).

[47] B. M. Jakosky and M. S. Team, presented at the Third International Workshop on The Mars Atmosphere: Modeling and Observations, Williamsburg, Virginia, 2008 (unpublished).